\documentclass[final,5p,times,twocolumn]{elsarticle}
\usepackage[utf8]{inputenc}
\usepackage{amsmath}
\usepackage{rotating}
\usepackage{graphicx}
\usepackage{axodraw4j}
\usepackage{pstricks}
\usepackage{color}
\def\qqbar{$q\overline q$ }

\def\ddbar{$d\overline d$ }
\def\uubar{$u\overline u$ }

\usepackage{lineno,hyperref}

\journal{Physics Letters B}

\begin{document}

\begin{frontmatter}

\title{{ Electroproduction ratios of Baryon-Meson states  and   Strangeness Suppression}}

\author[adress1]{ E. Santopinto\corref{mycorrespondingauthor}}
\cortext[mycorrespondingauthor]{Corresponding author}
\ead{santopinto@ge.infn.it}
\author[adress2]{R. Bijker} 
\author[adress1,adress2]{H. Garc{\'{\i}}a-Tecocoatzi}

 
\address[adress1]{INFN, Sezione di Genova, via Dodecaneso 33, 16146 Genova, Italy} 
\address[adress2]{Instituto de Ciencias Nucleares, Universidad Nacional Aut\'onoma de M\'exico, 04510 M\'exico DF, M\'exico}




\begin{abstract}
 We  describe the  electroproduction ratios of baryon-meson states from nucleon, inferring from the sea quarks in the
nucleon  using an extension of the quark model that takes into account the sea.  As a result  we provide, with no adjustable parameters,
 the predictions of ratios  of 
exclusive meson-baryon final states:$\Lambda K^+$, $\Sigma ^{*}K$, $\Sigma K$, $p\pi^0$, and $n\pi^+$. 
These predictions 
are in agreement with the new Jlab  experimental data showing  that sea quarks play an important role in  the  electroproduction. We also predicted  further  ratios of exclusive reactions that 
can be measured   and tested in future experiments.   In particular, we suggested new experiments on deuterium and tritium.
Such measurements can provide crucial test of different predictions concerning the structure of nucleon and its sea quarks helping to solve an outstanding problem. Finally, we computed the so called  strangeness suppression factor, $\lambda_s$, that is 
 the suppression of strange quark-antiquarks compared to nonstrange pairs,  and we found that  our finding with  this simple extension of the quark model is in good agreement  with the results of  Jlab and CERN experiments.
\end{abstract}

\begin{keyword}
Electroproduction\sep  Strangeness \sep Nucleon \sep arxiv:1601.06987
\MSC[2010] 00-01\sep  99-00
\end{keyword}

\end{frontmatter}


\section{Introduction}

In particle physics, the exact  hadronization process is unknown.
To determine the structure of  events at high and low energies a model for hadronization from first principles is needed but it is still missing. 
There have been many attempts at modeling \qqbar creation within the quark model (QM) formalism arising from Micu's suggestion
\cite{Micu:1968mk} that hadron decays proceed through $q \bar q$ pair production with vacuum quantum numbers, i.e. 
$J^{PC} = 0^{++}$. Since the $q \bar q$ pair corresponds to a $^3P_0$ quark-antiquark state, 
this model is known as the $^3P_0$ pair-creation model \cite{Micu:1968mk,LeYaouanc,Roberts:1992}.

New studies have been conducted by the CLAS Collaboration \cite{Mestayer:2014s}. These have  tried to extract  the flavor-dependence
of the \qqbar creation in  two-body exclusive reactions. The study of  \qqbar creation can help us to 
understand how  quarks become observable hadrons, which now is an open problem. 

In this letter, we will focus on the role of  nucleon sea quarks  in
 hadron production of baryon-meson final state for the exclusive reactions,
when there are no decay chains.  Since the   evidence for the flavor asymmetry of the proton sea was found 
by NMC at CERN \cite{nmc}, many studies have been carried out to explain the importance of the quark-antiquark content  in the nucleon and its 
role in the observables. 
We computed the ratios of  this baryon-meson production by inferring them from the  continiuum 
components of the nucleon using the Unquenched Quark Model (UQM)  for baryons\cite{Bijker:2012zza,Santopinto:2010zza,Bijker:2009up,BijSan:2010C}. We also extracted the  probabilities of the sea quarks and their relation with  \qqbar creation in electroproduction.  ~~~~~~~~~~~~~~~~~~~~~~~~~~~~~~~

\section{Unquenched Quark Model}
The    UQM
is based 
on a quark model (QM)  with  continuum components,  which  quark-antiquark pairs with vacuum quantum numbers are added as a 
perturbation employing a $^{3}P_0$ model for  $q \bar{q}$ pair creation.
This approach, which is a generalization of the unitarized quark model by T\"ornqvist and Zenczykowski \cite{Tornqvist}, was motivated by the work by Isgur and coworkers on the flux-tube breaking model. They showed that the QM emerges as the adiabatic limit of the flux-tube model to which the effects of \qqbar pair creation can be added as a perturbation \cite{Geiger:1996re}. 
 The pair-creation mechanism is inserted 
at the quark level and the one-loop diagrams are calculated by summing over a complete set of 
intermediate baryon-meson states. Under these assumptions, the baryon wave function consists 
of a zeroth order three-quark configuration $|A \rangle$ plus a sum over 
all the possible higher Fock components 
 due to the creation of $^{3}P_0$ quark-antiquark pairs.
\begin{multline}
|\psi_A \rangle ={\cal N}_a\left[ | A \rangle 
+\sum_{BC J_b J_c J_{bc} l} \int  k^2 dk \, 
| BC,(J_b J_c)J_{bc},l;J_a M_a;  k\rangle \right.\\ \ 
\left.\frac{ \langle BC, (J_b J_c)J_{bc},l;J_a M_a; k | T^{\dagger} | A \rangle } 
{m_a - E_b(k) - E_c(k)} \right] . \  \  \  \ 
\label{ucqmwf} 
\end{multline}
Here, $A$ denotes the initial baryon, $B$ and $C$ represent the asymptotic baryon and meson system,  $m_a$, $E_b(k)=\sqrt{m_b^2 + k^2}$ and $E_c(k)=\sqrt{m_c^2 + k^2}$ are their respective energies 
calculated in the rest frame of the initial baryon A, $\vec{k}$ and $l$ are the 
relative momentum and orbital angular momentum between $B$ and $C$ respectively, and $J_a$ is the 
total angular momentum $\vec{J}_a = \vec{J}_b + \vec{J}_c + \vec{l}$. The operator 
$T^{\dagger}$ creates a quark-antiquark pair in the $^{3}P_0$ state with the quantum 
numbers of the vacuum: $L=S=1$ and $J=0$
\begin{eqnarray}
T^{\dagger} &=& -3 \, \gamma \,\int d \vec{p}_4 \, d \vec{p}_5 \, 
\delta(\vec{p}_4 + \vec{p}_5) \, C_{45} \, F_{45} \,  
{e}^{-\alpha_d^2 (\vec{p}_4 - \vec{p}_5)^2/6 }\, 
\nonumber\\
&&  \left[ \chi_{45} \, \times \, {\cal Y}_{1}(\vec{p}_4 - \vec{p}_5) \right]^{(0)}_0 \, 
b_4^{\dagger}(\vec{p}_4) \, d_5^{\dagger}(\vec{p}_5) ~.   
\label{3p0}
\end{eqnarray}
Here, $b_4^{\dagger}(\vec{p}_4)$ and $d_5^{\dagger}(\vec{p}_5)$ are the creation operators for a quark 
and an antiquark with momenta $\vec{p}_4$ and $\vec{p}_5$, respectively. 
The quark-antiquark pair is characterized by a color singlet wave function $C_{45}$, a 
flavor singlet wave function $F_{45}$, a spin triplet wave function $\chi_{45}$ with spin 
$S=1$ and a solid spherical harmonic ${\cal Y}_{1}(\vec{p}_4 - \vec{p}_5)$ that indicates that 
the  quark and antiquark are in a relative $P$ wave. The operator $T^{\dagger}$ creates a pair 
of constituent quarks with an effective size. Thus, the pair creation point is smeared
out by a gaussian factor whose width is given by $\alpha_d$.

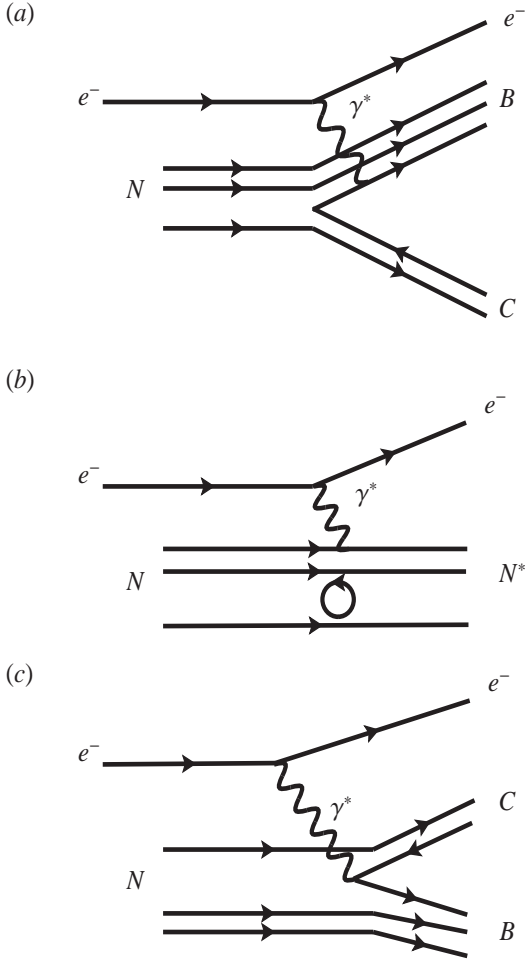
\begin{figure}[htbp] 
\caption{Schematic diagrams of  Electroproduction processes. The virtual baryon $B$ created by QCD vacuum  fluctuations absorbs the virtual photon (a). The virtual photon is absorbed by a constituent quark in the nucleon, thus a $N^*$ state is produced (b). The virtual photon creates a $q\bar{q}$ pair that can produce  a baryon-meson state (BC) from nucleon (c).  See the text for more details.}
\scalebox{0.45}[0.5]{
\fcolorbox{white}{white}{
 \begin{picture}(469,780) (30,-3)

 \label{Fig1} 
    \SetWidth{3.5}
    \SetColor{Black}
  \Line[arrow,arrowpos=0.5,arrowlength=10,arrowwidth=6,arrowinset=0.2](180,575)(304,575)
    \Line[arrow,arrowpos=0.5,arrowlength=10,arrowwidth=6,arrowinset=0.2](180,545)(304,545)
     \Line[arrow,arrowpos=0.5,arrowlength=10,arrowwidth=6,arrowinset=0.2](304,559)(448,623)
  \Line[arrow,arrowpos=0.5,arrowlength=10,arrowwidth=6,arrowinset=0.2](448,495)(304,560)
  \Line[arrow,arrowpos=0.5,arrowlength=10,arrowwidth=6,arrowinset=0.2](304,590)(448,655)
 \Line[arrow,arrowpos=0.5,arrowlength=10,arrowwidth=6,arrowinset=0.2](304,545)(448,479)
    \Line[arrow,arrowpos=0.5,arrowlength=10,arrowwidth=6,arrowinset=0.2](130,640)(304,640)
    \Line[arrow,arrowpos=0.5,arrowlength=10,arrowwidth=6,arrowinset=0.2](304,640)(448,700)
    \Photon(304,640)(350,580){7.5}{3}
    \Text(110,640)[lb]{\huge{\Black{$e^-$}}}
    \Text(464,700)[lb]{\huge{\Black{$e^-$}}}
    \Text(150,570)[lb]{\huge{\Black{$N$}}}
    \Text(460,639)[lb]{\huge{\Black{$B$}}}
    \Text(460,480)[lb]{\huge{\Black{$C$}}}
    \Line[arrow,arrowpos=0.5,arrowlength=10,arrowwidth=6,arrowinset=0.2](180,590)(304,590)
    \Line[arrow,arrowpos=0.5,arrowlength=10,arrowwidth=6,arrowinset=0.2](304,575)(448,639)
    \Text(336,630)[lb]{\huge{\Black{$\gamma^*$}}}
    \Line[arrow,arrowpos=0.5,arrowlength=10,arrowwidth=6,arrowinset=0.2](130,351)(304,351)
    \Line[arrow,arrowpos=0.5,arrowlength=10,arrowwidth=6,arrowinset=0.2](304,351)(431,399)
    \Line[arrow,arrowpos=0.5,arrowlength=10,arrowwidth=6,arrowinset=0.2](180,304)(432,304)
    \Line[arrow,arrowpos=0.5,arrowlength=10,arrowwidth=6,arrowinset=0.2](180,287)(431,287)
    \Arc[arrow,arrowpos=0.5,arrowlength=10,arrowwidth=6,arrowinset=0.2](325,266)(13,270,630)
    \Line[arrow,arrowpos=0.5,arrowlength=10,arrowwidth=6,arrowinset=0.2](180,246)(433,247)
    \Photon(305,351)(334,304){7.5}{3}
    \Text(110,355)[lb]{\huge{\Black{$e^-$}}}
    \Text(448,408)[lb]{\huge{\Black{$e^-$}}}
    \Text(50,700)[lb]{\huge{\Black{$(a)$}}}
    \Text(50,423)[lb]{\huge{\Black{$(b)$}}}
    \Text(150,274)[lb]{\huge{\Black{$N$}}}
    \Text(460,280)[lb]{\huge{\Black{$N^*$}}}
    \Text(50,201)[lb]{\huge{\Black{$(c)$}}}
    \Line[arrow,arrowpos=0.5,arrowlength=10,arrowwidth=6,arrowinset=0.2](130,142)(273,143)
    \Line[arrow,arrowpos=0.5,arrowlength=10,arrowwidth=6,arrowinset=0.2](273,143)(434,190)
    \Line[arrow,arrowpos=0.5,arrowlength=10,arrowwidth=6,arrowinset=0.2](180,78)(354,78)
    \Line[arrow,arrowpos=0.5,arrowlength=10,arrowwidth=6,arrowinset=0.2](354,78)(438,115)
    \Line[arrow,arrowpos=0.5,arrowlength=10,arrowwidth=6,arrowinset=0.2](436,98)(338,56)
    \Line[arrow,arrowpos=0.5,arrowlength=10,arrowwidth=6,arrowinset=0.2](339,55)(432,29)
    \Photon(272,142)(340,56){7.5}{5}
    \Line[arrow,arrowpos=0.5,arrowlength=10,arrowwidth=6,arrowinset=0.2](180,30)(354,30)
    \Line[arrow,arrowpos=0.5,arrowlength=10,arrowwidth=6,arrowinset=0.2](180,16)(354,16)
    \Line[arrow,arrowpos=0.5,arrowlength=10,arrowwidth=6,arrowinset=0.2](354,30)(432,16)
    \Line[arrow,arrowpos=0.5,arrowlength=10,arrowwidth=6,arrowinset=0.2](354,16)(432,-2)
    \Text(110,145)[lb]{\huge{\Black{$e^-$}}}
    \Text(451,198)[lb]{\huge{\Black{$e^-$}}}
    \Text(150,48)[lb]{\huge{\Black{$N$}}}
    \Text(460,108)[lb]{\huge{\Black{$C$}}}
    \Text(341,337)[lb]{\huge{\Black{$\gamma^*$}}}
    \Text(460,10)[lb]{\huge{\Black{$B$}}}
    \Text(320,100)[lb]{\huge{\Black{$\gamma^*$}}}
  \end{picture}
}
%
}
\end{figure}

\section{Production rates  in the UQM}

In the UQM, the asymptotic states are included in the wave function of the proton. These components are related to the creation of
$q\bar{q}$ pairs, with the quantum numbers  of the vacuum as it is described in the $^3P_0$ mechanism.
 Here we describe  the production 
ratios  from nucleon  $N\gamma^*\rightarrow BC/N\gamma^*\rightarrow B'C'$ in exclusive reactions.  The production ratios in exclusive reaction from proton target  were just measured by the CLAS Collaboration \cite{Mestayer:2014s}. We computed the ratios,  inferring from the ratio of probability to find the virtual-baryon-meson-components (BC)  in the nucleon wave function. 
\begin{eqnarray}
 P(BC)=   \left| \langle BC|\psi _{N}\rangle \right |^2
\end{eqnarray}

 In  the electroproduction process we could assume  that the proportional  constant could be  the same
 for baryon-meson final states, the production ratio from proton target,
for example,   could be written as follow:
\begin{eqnarray}
 \frac{p\rightarrow\Lambda K^+}{p\rightarrow n\pi^+}&\approx& \frac{P(\Lambda K^+)}{P(n\pi^+)}
 \label{relvr}
 \end{eqnarray}
{ [We have noticed that this assumption  is not always true, since in the process  a phase space factor should be added. However, in the recent study by Mestayer {\it et al.} \cite{Mestayer:2014s} the ratios between $W=1.65\  GeV$ and $W=2.55 \ GeV$ are approximately independent of $W$ (total hadronic energy in the center of mass frame). The complete theoretical dependence of the results on the energy is the subject of subsequent article \cite{future}.]

Eq. \ref{relvr} comes  from diagram (a) in the Fig. \ref{Fig1}, where the virtual photon $\gamma^*$ is absorbed  by the virtual baryon $B$, by the quark created in the  $^3P_0$ vertex. Under this picture, after the quark from the  $^3P_0$ vertex absorbs the photon, it can not close the $^3P_0$  loop and the hadronization occurs.   Since the  final baryon-meson state $BC$ is observed experimentally in exclusive reactions, only the  virtual component BC in the nucleon wave function will  contribute if $BC$  virtual state   is the same as the $BC$ state  that can be observed in exclusive reactions. In the ratios, the coupling  constants and other factors  due to the process cancel out, thus the ratios in exclusive reactions are inferred  from the sea quarks of the nucleon. 

Here we study the exclusive reactions where octet and decuplet  baryons in combination with pseudoscalar meson could be observed, thus the other components do  not contribute. To  extend our study to the description of baryon-vector meson production,   another process in diagram (a) (See Fig \ref{Fig1}) should be considered, where the photon  is absorbed by the antiquark in the meson C. The interference between this two diagrams would  be subject of subsequent article \cite{future}.   
The virtual photon can not couple to any valence quark, since if the virtual photon is absorbed by a valence quark  the $^3P_0$ loop would close, diagram (b) in the Fig. \ref{Fig1} , as a result a $N^*$ resonance is produced, that is not considered here because we are studying  the 
exclusive reactions. 

There exists another process in which the virtual photon creates a $q\bar q$ pair, that may contribute to the electroproduction of baryon-meson states, see Fig \ref{Fig1} diagram (c). The contribution of this creation process is expected to be small and is not considered here. 
   }
 
In the UQM the probability of finding a $BC$ component is   the product of a spin-flavor-isospin factor and a radial integral as follows:
\begin{eqnarray}
\frac{P(\Lambda K^+)}{P(n\pi^+)}
 =\frac{27}{50}
\frac{I_{N \rightarrow \Lambda K}}{I_{N \rightarrow N \pi}}~, 
\label{ratio}
\end{eqnarray}
where
the factor $ {27}/{50}$  is the ratio of the  color-flavor-spin-isospin (CFSI) factors and
 the integral $I$  is simply defined  in general for the production of a baryon B and a meson C, from a  baryon A,  as 
\begin{eqnarray*}
I_{A \rightarrow BC} &=& \int_{0}^{\infty} dk\frac{k^4 \mbox{e}^{-2F^2k^2}}{\Delta E_{A\rightarrow BC}^2(k)} ~. 
\end{eqnarray*}

The value of  $F^2$ depends on the size of the harmonic oscillator wave 
functions for baryons and mesons,  $\alpha_{\rm b}$, and  $\alpha_{\rm c}$ respectively,   and  the Gaussian smearing  of the pair-creation vertex, $\alpha_{d}$, it is taken from Ref. {\cite{Bijker:2009up}} to be  $  2.275$  GeV$^{-2}$.  In   Ref. {\cite{Bijker:2009up}},  $\alpha_{\rm b}$= 0.32 GeV, $\alpha_{\rm c}$= 0.40 GeV and  
$\alpha_{d}$= 0.30 fm.   
The energy denominator represents the energy difference between initial and final hadrons calculated in the 
rest frame of the initial baryon.

 The  color-flavor-spin-isospin factor plays 
an important role in the production rates. In  fact, for example  if we consider the 
limit case in which, the two final BC
states in the ratio  are  in the same isospin channels, the radial 
contribution is
the same and  the production rate will only depend on this factor. Thus the  predicted ratio for  $p \rightarrow p \pi^0 / p \rightarrow n \pi^+ = 1/2$ 
is well understood  on the basis of isospin symmetry (it is conserved in the UQM), for which the value is $1/2$. 
{However, the general result is determined by a combination of ratios of CFSI factors and integrals over the relative momentum, 
see Eq. ( \ref{ratio}). In the case of flavor symmetry, the ratios are determined completely by the CFSI factors.  
For example, in the limit of flavor symmetry the ratio $\Lambda K^+/n\pi^+$ would be ${27}/{50}=0.54$. The effect of $SU(3)$ symmetry breaking 
in the UQM can be observed by a comparison with UQM result $\Lambda K^+/n\pi^+=0.227$ from Table~\ref{tab:res}.     }
Inspection of Table~\ref{tab:res} shows that the observed rates are 
reproduced very well by our calculations. In addition, we present the results for some 
other channels. Moreover, using the isospin symmetry we computed the production rates from the neutron. We are aware that the neutron target does not exist but eventually those 
results can be extracted  from new experiments  on deuterium or tritium.    
\begin{table}[htbp]  
\caption{Comparison between    theoretical ratios of baryon-meson  electroproduction from neutron and proton   and experimental data. In the third column are
the theoretical results of the first and second columns. Experimental data is just for proton.}
\begin{center}
\begin{tabular}{cclc} 
\hline 
\hline \\
Ratio    &Ratio               &  UQM                 & Exp. \cite{Mestayer:2014s}  \\
from neutron&from proton&& from proton \\ \\
\hline \\
$\Lambda K^0/p\pi^-$&$\Lambda K^+/n\pi^+$    & 0.227                  & $0.19\pm0.01\pm0.03$ \\  
$\Lambda K^0/n\pi^0$&$\Lambda K^+/p\pi^0$    & 0.454                  & $0.50\pm0.02\pm0.12$ \\  
$n\pi^0/p\pi^-            $         &$p\pi^0/n\pi^+            $         & 0.5                   & $0.43\pm0.01\pm0.09$ \\
$\Sigma^0 K^0/p\pi^-$    &  $\Sigma^0 K^+/n\pi^+$    & 0.007                  & $-$ \\ 
$\Sigma^0 K^0/n\pi^0$    &$\Sigma^0 K^+/p\pi^0$    & 0.014                  & $-$ \\ 
$\Sigma^- K^+/p\pi^-$    & $\Sigma^+ K^0/n\pi^+$    & 0.014                  & $-$ \\ 
$\Sigma^- K^+/n\pi^0$    &   $\Sigma^+ K^0/p\pi^0$    & 0.028                  & $-$ \\
$\Sigma^{*0} K^0/p\pi^-$    & $\Sigma^{*0} K^+/n\pi^+$    & 0.045                  & $-$ \\ 
$\Sigma^{*0} K^0/n\pi^0$ & $\Sigma^{*0} K^+/p\pi^0$    & 0.090                  & $-$ \\ 
$\Sigma^{*-} K^+/p\pi^-$    & $\Sigma^{*+} K^0/n\pi^+$    & 0.090                  & $-$ \\ 
$\Sigma^{*-} K^+/n\pi^0$    & $\Sigma^{*+} K^0/p\pi^0$    & 0.18                  & $-$ \\ 
\hline 
\hline
\end{tabular}
\end{center}
\label{tab:res}  
\end{table}

\section{Extraction of \qqbar-creation-probability and the strangeness suppression factor}

 As we have already shown, the asymptotic states play an important  role in exclusive two-body production, since 
  the  ratios of this process can be determined in 
a straightforward way by means of the UQM inferred from the virtual components produced by the quantum  vacuum fluctuations. Likewise  we can infer from the flavor content of the proton the 
\qqbar creation probabilities on the $BC$ final states.  Within this approach the production of the \qqbar in   the asymptotic states of the proton is linked with 
the final states  produced by electroproduction.
 Here, we extract the probabilities from  sea quarks  and  we only point out  the   fact  that the production of \qqbar is related to the sea quarks in the 
nucleon. As an example we study  the strangeness suppression 
factor (SSF) $\lambda_s$ is defined  in the literature \cite{Bocquet:1996} as:
 
 \begin{eqnarray}\lambda_s=\frac{2 (s\bar{s})}{(u\bar{u})+(d\bar{d})}
 \end{eqnarray}
 { 
 where $u\bar{u}$, $d\bar{d}$ and  $s\bar{s}$ are the mean numbers of $u,d, s$ quarks and the corresponding antiquarks in the final state. The probabilities  to produce a $q\bar {q}$ pair  are computed
 \begin{eqnarray}
{ q\bar{q}}={\sum_{B_{i}C{i}}\langle B_{i}C_{i}|\hat{P}(q\bar{q})|N\rangle^2}
\end{eqnarray}  
where  $\hat{P}(q\bar{q})$ is the operator that counts the $q\bar {q}$ pairs created in the electroproduction  from nucleon, $B_{i}$ and $C_{i}$ are the all possible  baryon and meson states that can be observed in exclusive reactions. 
} 


{
Our results  for the \qqbar production ratios are present in   Table~\ref{strange}, 
 the column UQM$^{(1)}$ contains   the \qqbar production ratios  with  pseudoscalar  mesons in combination  with octet and decuplet baryons ($N\pi$, $\Delta\pi$, $\Sigma K$, $\Sigma^* K$, $\Lambda K $, $N\eta $, $N \eta'$ channels), while the column  UQM$^{(2)}$  only takes into account $\pi$ and $K$ mesons in combination  with octet and decuplet baryons ($N\pi$, $\Delta\pi$, $\Sigma K$, $\Sigma^* K$, $\Lambda K $ channels). 
 
Since the isospin symmetry  is valid for  the nucleon  wave function in the   UQM, we present  our results   for the \qqbar production ratios  of  proton and  neutron  in Table\ref{strange}. }

The value for the strangeness suppression factor 
$\lambda_s$ is in good agreement with the observed values from exclusive reactions  at JLAB 
\cite{Mestayer:2014s} and in high-energy production  experiments   at CERN \cite{Bocquet:1996}.

\begin{table}[t]
\begin{center}
\caption{The production ratios of the \qqbar  probabilities and $\lambda_s$  inferred
from sea quarks of the nucleon. The column UQM$^{(1)}$ contains   the \qqbar production ratios  with  $\pi$, $K$, $\eta$  and $\eta'$ mesons, while the column  UQM$^{(2)}$  only takes into account $\pi$ and $K$ mesons.}
\label{strange}
\begin{tabular}{c}
\begin{tabular}{ccccc}
\hline
\hline
&Proton&&&\\
\hline
\noalign{\smallskip}
Ratio                           & UQM$^{(1)}$  &UQM$^{(2)}$& Exp. & Ref. \\
\noalign{\smallskip}
\hline
\noalign{\smallskip}
$s\bar{s}/d\bar{d}$             & $0.265$ &$0.245$& $0.22 \pm 0.07$ & \cite{Mestayer:2014s} \\
$u\bar{u}/d\bar{d}$             & $0.568$ &$0.568$& $0.74 \pm 0.18$ & \cite{Mestayer:2014s} \\ 
$2s\bar{s}/(u\bar{u}+d\bar{d})$ & $0.338$ &$0.313$& $0.25 \pm 0.09$ & \cite{Mestayer:2014s} \\ 
                                &         && $0.29 \pm 0.02$ & \cite{Bocquet:1996} \\
\noalign{\smallskip}
\hline
&Neutron&&&\\
\hline
\\

\noalign{\smallskip}
$s\bar{s}/u\bar{u}$             & $0.265$ &$0.245$&  \\
$d\bar{d}/u\bar{u}$             & $0.568$ &$0.568$&  &\\ 
$2s\bar{s}/(u\bar{u}+d\bar{d})$ & $0.338$ &$0.313$& &  \\ 
                            
\noalign{\smallskip}
\hline
\hline
\end{tabular}
\end{tabular}
\end{center} 
\end{table}

\section{Summary and conclusions}

In summary we computed the production  ratios of baryon-meson final states in exclusive reactions and we  found that our  predictions  of those  ratios are in good agreement with the new experimental data at Jlab \cite{Mestayer:2014s} as a first theoretical  calculation that explains  the new data reported in the letter
\cite{Mestayer:2014s}.
 We also computed  other  baryon-meson production ratios  still unobserved  but   that in principle could be  observed  in exclusive reactions from proton or neutron initial states in future  new experiments. In particular,  for the neutron case   we suggested new experiments  on deuterium or tritium. We showed that sea quarks play an important role  in the electroproduction  since our predictions are in agreement with the experimental results, thus 
we found that  the meson-baryon production 
in exclusive reactions   could provide information  on the sea components of the  nucleon.  
Therefore future  experiments of exclusive reactions can provide another test of different predictions concerning the nucleon structure. 


An interesting result was the production ratio $u\bar{u}/d\bar{d}=0.568$ from the proton, which is compatible with the experimental value obtained by Mestayer {\it et al}  at JLAB  \cite{Mestayer:2014s}, unlike  the unity value used in high energy hadronization models. Those two finding reinforce  each other ( the experimental one by   \cite{Mestayer:2014s}, and our theoretical prediction ) and in our  view  are  well explained  from  the fact that there is  a  \ddbar excess  on proton as  seen by NMC and HERMES  \cite{nmc,hermes}  and if one believes  that  the \qqbar production is governed by the sea quark on proton, where the  \ddbar is bigger than the \uubar content, this result on the ratio different from one is a  direct consequence.  In the case of neutron,  we predicted   $d\bar{d}/u\bar{u}=0.568$ thus  the \uubar production would be higher than the \ddbar production, since isospin symmetry yields a  \uubar  excess  over \ddbar in neutron, but this prediction can be tested by future experiments at Jlab and LHC.  On the contrary it is worthwhile noticing  that $\lambda_s$ is the same from proton and neutron.


The extraction of quark-antiquark production  in exclusive reactions from nucleon gave us a  suppression of strange quark-antiquarks compared to nonstrange pairs.  This  is in a  good agreement with the available experimental data both  from JLAB \cite{Mestayer:2014s} and CERN \cite{Bocquet:1996}. 

Finally, in ref. \cite{Mestayer:2014s}  the experimentalists observed that the measured rates are independent from the type and energy of any resonance formed in the intermediate state. As such, the results should reflect some fundamental aspect of flavor production like the breaking of the underling  SU(3) symmetry  induced by the sea quark components. { Moreover, our picture can be extended to  study the photoproduction and  hadron production with meson beams.  It will  be the   subject of a 
future article \cite{future}, that eventually can be tested at Jpark.}

It is worthwhile  noticing  that the results for the UQM are independent of the overall parameter or  strength of the $^{3}P_0$ quark-antiquark 
pair creation vertex, usually called $\gamma_0$. The values of the remaining parameters were taken from previous work and no attempt 
was made to optimize their values. 

\section*{Acknowledgments}
The authors acknowledge  interesting discussion with  Dr. Daniel  Carman ( Jlab).
This work is supported in part by INFN sezione di Genova and PAPIIT-DGAPA, Mexico (grant IN107314).

\section*{References}



\end{document}